\documentclass[hyper]{JHEP} 
\usepackage{amsmath,amssymb,amsfonts}

\usepackage{epsfig}

\newcommand\fverb{\setbox\pippobox=\hbox\bgroup\verb}

\newcommand\fverbdo{\egroup\medskip\noindent%

            \fbox{\unhbox\pippobox}\ }

\newcommand\fverbit{\egroup\item[\fbox{\unhbox\pippobox}]}

\newbox\pippobox

\title{Hamiltonian Analysis of Minimal  Massive Gravity Coupled
to Galileon Tadpole Term}
\author{J. Kluso\v{n}\\
Department of
Theoretical Physics and Astrophysics\\
Faculty of Science, Masaryk University\\
Kotl\'{a}\v{r}sk\'{a} 2, 611 37, Brno\\
Czech Republic\\
E-mail: \email{klu@physics.muni.cz}}
\preprint{}

 \abstract{We perform the Hamiltonian analysis of minimal massive gravity
 coupled to the Galileon tadpole term. We determine all   constraints
  and we argue that
  the physical degrees of freedom correspond to ten modes of the massive
 gravity together with $2(D-3)$ Galileons so that given model is ghost  free.}

\def\bA{\mathbf{A}}

\def\hf{\hat{f}}

\def\bSigma{\mathbf{\Sigma}}

\def\mC{\mathcal{C}}
\def\be{\begin{equation}}

\def\bC{\mathbf{C}}
\def\ee{\end{equation}}

\def\tn{\tilde{n}}
\def\tx{\tilde{x}}
\def\tK{\tilde{K}}
\def\mR{\mathcal{R}}
\def\mM{\mathcal{M}}
\def\tSigma{\tilde{\Sigma}}
\def\bea{\begin{eqnarray}}

\def\eea{\end{eqnarray}}

\def\mH{\mathcal{H}}

\def\bx{\mathbf{x}}
\def\by{\mathbf{y}}

\newcommand{\hg}{\hat{g}}

\newcommand{\tD}{\tilde{D}}
\newcommand{\mG}{\mathcal{G}}

\def \bA{\mathbf{A}}

\newcommand{\bT}{\mathbf{T}}

\def\pb #1{\left\{#1\right\}}

\begin{document}
\section{Introduction and Summary}\label{first}
The non-linear massive gravity is very nice and intriguing proposal
of the ghost-free massive gravity (dRGT)
\cite{deRham:2010kj,Hassan:2011hr}. dRGT massive gravity  is
$3-$parameter of potentials whose structure is based on the square
root of the matrix $\hg^{\mu\nu}\hf_{\nu\sigma}$ where
$\hg_{\mu\nu}$ is dynamical four dimensional metric while
$\hf_{\mu\nu}$ is fixed four dimensional metric. The structure of
given potential ensures an existence of the additional constraint
that is sufficient for removing the ghost degree of freedom. Further
important extension of given work was performed in
\cite{Hassan:2011vm,Hassan:2011hr,Hassan:2011tf,Hassan:2011zd} while
the general proof of the absence of the ghosts using the Hamiltonian
formalism was presented in \cite{Hassan:2011ea}, for further works,
see
\cite{Golovnev:2011aa,Kluson:2011rt,Kluson:2012gz,Kluson:2012wf,Kluson:2012zz,
Comelli:2012vz,Comelli:2013paa,Comelli:2013txa}.

Generally the presence of the mass term with fixed metric
$\hf_{\mu\nu}$ breaks the diffeomorphism invariance completely,
however this can be restored by introduction of four the
St\"{u}ckelberg fields so that $\hf_{\mu\nu} \Rightarrow
\partial_\mu
\phi^A\partial_\nu\phi^B\eta_{AB}$ \cite{deRham:2011rn}
\footnote{For Hamiltonian analysis of the dRGT gravity formulated
with St\"{u}ckelberg fields, see \cite{Hassan:2012qv} and also
\cite{Kluson:2011rt,Kluson:2012gz,Kluson:2012wf,Kluson:2012zz}.}. By
construction St\"{u}ckelberg fields are pure gauge and original dRGT
theory can be restored by gauge fixing of the diffeomorphism
invariance by imposing the conditions $\phi^A=\delta^A_\mu x^\mu$.

The generalization of given construction was presented recently in
\cite{Gabadadze:2012tr,Andrews:2013ora} with following  basic idea:
Let us interpret the St\"{u}ckelberg fields as the embedding mapping
of a sigma model $\Sigma\rightarrow \mathcal{M}$ where both $\Sigma$
and $\mathcal{M}$ are four dimensional Minkowski space-time
\footnote{For related works, see \cite{Lin:2013aha}.}. In this
picture the dynamical metric $\hg_{\mu\nu}(x)$ is a world-volume
metric that lives on $\Sigma$. Then there is a natural
generalization when we consider the target space to be
higher-dimensional and also it can be curved. In other words we can
consider the metric $\hf_{\mu\nu}$ in the form
\begin{equation}
\hf_{\mu\nu}=\partial_\mu\phi^A\partial_\nu\phi^B\mG_{AB}(\phi) \ ,
\end{equation}
where now $A,B,C...=0,\dots,D$. Due to this generalization we now
have $D-3$ scalar fields that cannot be gauged away so that they are
physical scalars that couple to the physical metric through the dRGT
potential \cite{Gabadadze:2012tr}. Due to the fact that given fields
are dynamical we can consider more general form of the action that
contain invariants constructed solely from $\hf_{\mu\nu}$. The
leading term in this construction is the DBI action (known as the
tadpole term) $\sim \int d^4x\sqrt{-\det\hf_{\mu\nu}}$ and the
higher Lovelock invariants give Galileons
\cite{deRham:2010eu,Hinterbichler:2010xn}. It turns out that this
theory possesses a Galileon-like symmetry for each isometry.

The construction presented in \cite{Gabadadze:2012tr}  is very
interesting from different points of views. It is a generalization
of the dRGT theory but also provides a way how to couple the
Galileons to massive gravity while preserving the Galileon
invariance. This is very important fact since it is well known that
when we try to couple the Galileon to massless gravity we have to
introduce non-minimal coupling in order to ensure the second order
equations of motion and the Galileon symmetry is broken
\cite{Deffayet:2009wt,Deffayet:2009mn}. These results suggest that
Galileon could couple more naturally to the massive gravity than to
the ordinary massless gravity.

It was argued  in \cite{Gabadadze:2012tr} that given theory is
ghost-free for the flat target space metric in the decoupling limit
and for simplifying choice of parameters. Then it was argued in
\cite{Andrews:2013ora}, using the methods similar to those
\cite{Hinterbichler:2012cn} that the full theory, for any target
space metric $\mG_{AB}$ has the primary constraints that is
necessary to eliminate the Boulware-Deser ghost.

Despite of this remarkable conclusion  we fell that the coupled
system of massive gravity and Galileon deserves further
investigation  from different  point of view. Explicitly,we are not
quit sure whether  the  analysis presented in \cite{Andrews:2013ora}
is sufficient to show that the ghost mode is eliminated since  even
if they identified the constraints that could eliminate the ghost
mode it was not shown whether they are the first or the second class
constraints. Further, it is not clear how to find the momenta
conjugate to the scalar modes since the coupled action between
massive gravity and Galileon is rather complicated and it is not
immediately clear how to perform the Legendre transform from
Lagrangian to Hamiltonian formulation. Our goal is not to perform
the analysis of the  dRGT theory coupled to the Galileon actions in
full generality. Rather we will be more modest and perform the
Hamiltonian analysis of the particular model of the minimal dRGT
gravity coupled with  Galileon tadpole term that allows an explicit
analysis. It turns out that it is convenient  to perform this
analysis when we consider  dRGT theory with redefined shift
functions
\cite{Hassan:2011vm,Hassan:2011hr,Hassan:2011tf,Hassan:2011zd}. Then
we will be able to perform the Hamiltonian analysis and find
corresponding primary constraints. The analysis is similar to the
analysis performed in case of pure massive gravity in
\cite{Kluson:2012zz} however now the presence of the Galileon
tadpole term makes it more complicated. Despite of this fact we find
the constraint structure of given theory and  determine the
character of given constraints. We show that there are really two
additional constraints whose presence allow to eliminate the ghost
mode. In other words  our result confirms the results presented in
\cite{Andrews:2013ora} using the metric formulation of massive
gravity at least for some particular case of minimal dRGT theory
coupled to tadpole Galileon term.

The work presented here can be extended in different way. In
particular, we could consider the general form of dRGT theory
coupled  to the Galileon or the minimal dRGT theory  coupled to the
general Galileon Lagrangian or finally the most general case of the
general dRGT massive gravity  coupled to the general Galileon
action. However in all these cases the analysis is very complicated.
In particular, the simplest analysis could be in case of general
dRGT gravity coupled with the Galileon tadpole term where when we
can follow \cite{Kluson:2012zz} and determine all constraints of the
theory. On the other hand it is very difficult to determine the
character of these constraints in case of the general dRGT massive
gravity  due to their complicated form as was shown in
\cite{Kluson:2012zz}. The situation could be even worse in case of
more general Galileon term since it seems to be very difficult to
express momenta as function of the time derivatives of the scalar
fields and hence to perform the Legendre transformation from the
Lagrangian to the Hamiltonian formalism.

This note is organized as follows. In the next section
(\ref{second}) we introduce the minimal version of dRGT gravity
coupled to the Galileon tadpole term formulated with the transformed
shift function. Then in section (\ref{third}) we perform the
Hamiltonian analysis of given action and argue that the constraint
structure of given theory allows to eliminate the ghost mode.
Finally in Appendix we briefly review the Hamiltonian analysis of
the Galileon tadpole term.

\section{Minimal dRGT Massive Gravity  Coupled with  Galileon Tadpole
Term}\label{second} Let us consider minimal dRGT theory  coupled
with the tadpole Galileon  term. The action of the system has the
form
\begin{eqnarray}
S&=&S_{m.g.}+S_{gal} \ , \nonumber \\
 S_{m.g.}&=&M_p^2 \int d^4x\sqrt{-\hg}
\left[{}^{(4)}R[\hg]+2m^2(3-\sqrt{\hg^{-1}\hf})\right] \ , \nonumber \\
\end{eqnarray}
 where
\begin{equation}\label{Sgalltad}
S_{gal}=-T\int d^4x\sqrt{-\det \hf_{\mu\nu}} \ ,
\end{equation}
and where $\hg_{\mu\nu}$ is four dimensional metric with signature
$(-,+,+,+)$, ${}^{(4)}R[\hg]$ is scalar curvature calculated with
$\hg_{\mu\nu}$ and finally $\hf_{\mu\nu}$ is induced metric on the
world-volume of brane defined as
$\hf_{\mu\nu}=\partial_\mu\phi^A\mG_{AB}\partial_\nu\phi^B$. For
simplicity of further analysis we consider the case when $\mG_{AB}$
is the constant tensor keeping in mind that the generalization to
the case when $\mG_{AB}$ depends on $\phi$ is straightforward.
Finally, $M_p$ is four dimensional Planck mass and $T$ is the
tension of four-dimensional brane.

The coupling between gravity and Galileon is described through the
massive term $\sqrt{\hg^{\mu\nu}\hf_{\nu\rho}}$ where the square
root is defined as $\sqrt{\hg^{\mu\nu}\hf_{\nu\rho}}
\sqrt{\hg^{\rho\sigma}\hf_{\sigma\omega}}=
\hg^{\mu\nu}\hf_{\nu\omega}$. To proceed further we use  $3+1$
decomposition of the four dimensional metric $\hg_{\mu\nu}$
\cite{Gourgoulhon:2007ue,Arnowitt:1962hi}
\begin{eqnarray}
\hat{g}_{00}=-N^2+N_i g^{ij}N_j \ , \quad \hat{g}_{0i}=N_i \ , \quad
\hat{g}_{ij}=g_{ij} \ ,
\nonumber \\
\hat{g}^{00}=-\frac{1}{N^2} \ , \quad \hat{g}^{0i}=\frac{N^i}{N^2} \
, \quad \hat{g}^{ij}=g^{ij}-\frac{N^i N^j}{N^2} \
\nonumber \\
\end{eqnarray}
so that we find
\begin{equation}
N^2\hg^{-1}f= \left(\begin{array}{cc} -f_{00}+N^l f_{l0} &
-f_{0j}+N^l f_{lj}
\\
N^2 g^{il}f_{l0}-N^i(-f_{00}+N^l f_{l0}) & N^2
g^{il}f_{lj}-N^i(-f_{0j}+ N^l f_{lj}) \\ \end{array}\right) \ ,
\end{equation}
Following
\cite{Hassan:2011vm,Hassan:2011hr,Hassan:2011tf,Hassan:2011zd} we
perform the redefinition of the shift function $N^i$
\begin{equation}\label{defNi}
N^i=M\tn^i+f^{ik}f_{0k}+N \tD^i_{ \ j } \tn^j \ ,
\end{equation}
where
\begin{equation}\label{deftx}
 \tx=1-\tn^i f_{ij}\tn^j \ , \quad
 M^2=-f_{00}+f_{0k}f^{kl}f_{l0}
\end{equation}
and where we defined $f^{ij}$ as the inverse to $f_{ij}$ in the
sense \footnote{Note that in our convention $f^{ik}$ coincides with
$({}^{3}f^{-1})^{ik}$ presented in
\cite{Hassan:2011zd,Hassan:2011tf,Hassan:2011hr,Hassan:2011vm}.}
\begin{equation}
f_{ik}f^{kj}=\delta_i^{ \ j} \ .
\end{equation}

Finally note that the matrix
 $\tD^i_{ \ j}$ obeys the equation
 \cite{Hassan:2011vm,Hassan:2011hr,Hassan:2011tf,Hassan:2011zd}
\begin{eqnarray}\label{deftD}
\sqrt{\tx}\tD^i_{ \ j}= \sqrt{(g^{ik}-\tD^i_{ \ m} \tn^m
\tD^k_{ \ n}\tn^n)f_{kj}} \  \nonumber \\
\end{eqnarray}
and also following important identity
\begin{eqnarray}
f_{ik}\tD^k_{ \ j}= f_{jk}
\tD^k_{ \ i } \ .  \nonumber \\
\end{eqnarray}
Now we proceed to the case of the tadpole Galileon action. Using the
property of the determinant we obtain
\begin{eqnarray}\label{Sgall2}
S_{gal}&=& -T\int d^4x \sqrt{-\det f_{\mu\nu}}=\nonumber \\
&=&
 -T \int d^4x
\sqrt{-(f_{00}-f_{0i}f^{ij}f_{j0})}\sqrt{\det f_{ij}}=
-T\int d^4x M\sqrt{f} \ ,  \nonumber \\
\end{eqnarray}
where
\begin{equation}
 f\equiv \det f_{ij} \ .
\end{equation}
 Then using the results derived in
\cite{Hassan:2011vm,Hassan:2011hr,Hassan:2011tf,Hassan:2011zd} and
(\ref{Sgall2}) we find the action in the form
\begin{eqnarray}\label{massgr2}
S=M_p^2\int d^3\bx dt [N\sqrt{g}
\tK_{ij}\mG^{ijkl}\tK_{kl}+N\sqrt{g}R -\sqrt{g}MU'
-2m^2(N\sqrt{g}\sqrt{\tx}D^i_{ \ i}-3N\sqrt{g})] \ ,
\nonumber \\
\end{eqnarray}
where
\begin{equation}
U'=2m^2\sqrt{\tx}+\frac{T}{M_{p}^2}\frac{\sqrt{f}}{\sqrt{g}}
 \ ,
 \end{equation}
 and
  where we used the $3+1$ decomposition of the four dimensional
scalar curvature
\begin{equation}\label{31R}
{}^{(4)}R=\tK_{ij}\mG^{ijkl}\tK_{kl}+R \ ,
\end{equation}
where $R$ is three dimensional scalar curvature and where
\begin{equation}
\mG^{ijkl}=\frac{1}{2}(g^{ik}g^{jl}+g^{il}g^{jk})- g^{ij}g^{kl}
\end{equation}
with inverse
\begin{equation}
\mG_{ijkl}=\frac{1}{2}(g_{ik}g_{jl}+g_{il}g_{jk})-\frac{1}{2}g_{ij}g_{kl}
\ , \quad
\mG_{ijkl}\mG^{klmn}=\frac{1}{2}(\delta_i^m\delta_j^n+\delta_i^n\delta_j^m)
\ .
\end{equation}
Note that  in (\ref{31R}) we  ignored the total derivative terms.
Finally note that $\tK_{ij}$ is defined as
\begin{equation}
\tK_{ij}=\frac{1}{2N}(\partial_t g_{ij}- \nabla_i
N_j(\tn,g)-\nabla_j N_i(\tn,g)) \ ,
\end{equation}
where $N_i$ depends on $\tn^i$ and $g$ through the relation
(\ref{defNi}).
\section{Hamiltonian Formalism}\label{third}
Now we are ready to proceed to the Hamiltonian formalism, following
\cite{Kluson:2012zz}. From (\ref{massgr2}) we find the momenta
conjugate to $N,\tn^i$ and $g_{ij}$
\begin{equation}
\pi_N\approx 0 \ , \quad  \pi_i\approx 0 \ , \quad
\pi^{ij}=M_p^2\sqrt{g}\mG^{ijkl}\tK_{kl} \
\end{equation}
and the momentum conjugate to $\phi^A$
\begin{eqnarray}
p_A
=- \left(\frac{\delta M}{\delta\partial_t \phi^A}
\tn^i+f^{ij}\partial_j\phi_A\right) \mR_i-M_p^2\sqrt{g} \frac{\delta
M}{\partial_t \phi^A}
U' \ , \nonumber \\
\end{eqnarray}
where
\begin{equation}
 \mR_i=-2g_{ik}\nabla_j\pi^{kj} \ .
\end{equation}
 It turns out that it is useful to write $M^2$ in the form
\begin{eqnarray}
M^2=-\partial_t\phi^A \mM_{AB}\partial_t \phi^B \ , \quad
\mM_{AB}=\eta_{AB}-\partial_i\phi_A f^{ij}\partial_j\phi_B \ ,
\nonumber \\
\end{eqnarray}
where by definition the matrix $\mM_{AB}$ obeys following relations
\begin{eqnarray}\label{mMprop}
\mM_{AB}\eta^{BC}\mM_{CD}=
\mM_{AD} \ , \quad  \det \mM^A_{ \ B}=1
\  \nonumber \\
\end{eqnarray}
together with
\begin{eqnarray}\label{partiphimM}
\partial_i\phi^A \mM_{AB}=
\partial_i\phi_B-
\partial_i\phi^A\partial_k\phi_A
f^{kl}\partial_l\phi_B=0 \ .
\end{eqnarray}
With the help of these results we find
\begin{eqnarray}\label{PiAM}
p_A+\mR_if^{ij}\partial_j\phi_A= (\tn^i\mR_i+M_p^2\sqrt{g}U')
\frac{1}{M}\mM_{AB}\partial_t \phi^B \
\end{eqnarray}
and then following primary constraint
\begin{equation}\label{defSigmapo}
 \Sigma_p=(\tn^i
\mR_i+M_p^2\sqrt{g}U')^2+(p_A+\mR_if^{ij}\partial_j\phi_A)
(p^A+\mR_i f^{ij}\partial_j\phi^A)\approx
0 \ . \nonumber \\
\end{equation}
Note that  using (\ref{partiphimM}) we obtain another set of the
 primary constraints
\begin{eqnarray}\label{defSigmai}
\partial_i\phi^A\Pi_A=
\partial_i\phi^A
p_A+\mR_i=\Sigma_i\approx  0 \ .  \nonumber
\\
\end{eqnarray}
Observe that using (\ref{defSigmai}) we can write
\begin{eqnarray}
p_A+\mR_if^{ij}\partial_j\phi_A=\mM_{AC}\eta^{CB}p_B+\Sigma_i
f^{ij}\partial_j \phi_A
\end{eqnarray}
so that we can rewrite $\Sigma_p$ into the form
\begin{equation}
\Sigma_p= (\tn^i \mR_i+M_p^2\sqrt{g}U')^2+p_A\mM^{AB}p_B+H^i\Sigma_i
 \  \ ,
\end{equation}
where $H^i$ are  functions of the phase space variables. As a result
we see that it is natural to consider following independent
constraint $\Sigma_p$
\begin{equation}\label{defSigmap}
\Sigma_p= (\tn^i \mR_i+M_p^2\sqrt{g}U')^2+p_A\mM^{AB}p_B\approx 0 \
.
\end{equation}
We return to the analysis of the constraint $\Sigma_p$ below.

Now we are ready to write the extended Hamiltonian which includes
all the primary constraints
\begin{equation}
H_E=\int d^3\bx (N\mC_0+v_N\pi_N+v^i\pi_i+\Omega_p\Sigma_p+
\Omega^i\tSigma_i) \ ,
\end{equation}
where
\begin{eqnarray}
\mC_0= \frac{1}{\sqrt{g}M_p^2} \pi^{ij}\mG_{ijkl}\pi^{kl}-M_p^2
\sqrt{g} R+ 2m^2M_p^2\sqrt{g}\sqrt{\tx}\tD^i_{ \ i}
-6m^2M_p^2\sqrt{g}+
\tD^i_{ \ j}\tn^j\mR_i \nonumber \\
\end{eqnarray}
and where we introduced the constraints $\tSigma_i$ defined as
\begin{equation}
\tSigma_i=\Sigma_i+\partial_i \tn^i\pi_i+
\partial_j(\tn^j\pi_i) \ .
\end{equation}
Note that $\tSigma_i$ is defined as    linear combination of the
constraints $\Sigma_i\approx 0$ together with  the constraints
$\pi_i\approx 0$.

To proceed further we have to check the stability of all
constraints. To do this  we have to calculate the Poisson brackets
between all constraints and the Hamiltonian $H_E$. Note that  we
have following set of the canonical variables
$g_{ij},\pi^{ij},\phi^A,p_A,\tn^i,\pi_i$ and $N,\pi_N$ with
 non-zero Poisson brackets
\begin{eqnarray}\label{defpb}
\pb{g_{ij}(\bx),\pi^{kl}(\by)}&=& \frac{1}{2} (\delta_i^k\delta_j^l+
\delta_i^l\delta_j^k)\delta(\bx-\by) \ , \quad
\pb{\phi^A(\bx),p_B(\by)}=\delta^A_B\delta(\bx-\by) \ , \nonumber \\
\pb{N(\bx),\pi_N(\by)}&=&\delta(\bx-\by) \ , \quad
\pb{\tn^i(\bx),\pi_j(\by)}=\delta^i_j\delta(\bx-\by) \ . \nonumber
\\
\end{eqnarray}
The constraint $\tSigma_i$ has the same form as in
\cite{Kluson:2012zz} where it was shown that
 the smeared form of this constraint
\begin{equation}\label{defbTS}
 \bT_S(N^i)=\int d^3\bx
N^i\tSigma_i \
\end{equation}
 is the generator of the spatial diffeomorphism so that
\begin{eqnarray}
\pb{\bT_S(N^i),\tn^k}
&=&-N^i\partial_i\tn^k+\tn^j\partial_jN ^k \ ,  \nonumber \\
\pb{\bT_S(N^i),\mR_j}&=&-\partial_i N^i \mR_j
-N^i\partial_i\mR_j-\mR_i\partial_j N^i \ ,
 \nonumber \\
\pb{\bT_S(N^i),p_A}&=&-N^i\partial_ip_A-
\partial_i N^ip_A \ , \nonumber \\
\pb{\bT_S(N^i),\phi^A}&=&-N^i\partial_i\phi^A \ ,
\nonumber \\
\pb{\bT_S(N^i),g_{ij}}&=& -N^k\partial_k g_{ij}-\partial_i
N^k g_{kj}-g_{ik}\partial_j N^k \ , \nonumber \\
\pb{\bT_S(N^i),\pi^{ij}}&=& -\partial_k (N^k \pi^{ij}) +
\partial_k N^i\pi^{kj}+\pi^{ik}\partial_k N^j \ , \nonumber \\
\pb{\bT_S(N^i),f_{ij}}&=&-N^k\partial_k f_{ij}-\partial_i N^k
f_{kj}-f_{ik}
\partial_j N^k \ , \nonumber \\
\pb{\bT_S(N^i),\pi^i}&=& -\partial_i N^i\pi^j -N^i\partial_i
\pi^j+\partial_j N^i\pi^j \
 \nonumber \\
 \end{eqnarray}
and also
\begin{eqnarray}\label{pbbTSmC0}
\pb{\bT_S(N^i),\mC_0}&=&-N^m\partial_m \mC_0-
\partial_m N^m \mC_0 \ , \nonumber \\
 \pb{\bT_S(N^i),\Sigma_p}&=&-N^m\partial_m \Sigma_p
 -\partial_m N^m\Sigma_p \ .  \nonumber \\
\end{eqnarray}
Finally  it is easy to show that following Poisson bracket holds
\begin{equation}\label{bTSNM}
\pb{\bT_S(N^i),\bT_S(M^j)}= \bT_S(N^j\partial_j M^i-M^j
\partial_j N^i) \ .
\end{equation}
Now we are ready to analyze the stability of all
 primary
constraints. As usual the
 requirement of the preservation of the
constraint $\pi_N\approx 0$ implies an existence of the secondary
constraint $\mC_0\approx 0$. However the fact that $\mC_0$ is the
constraint immediately implies that  the constraint
$\tSigma_i\approx 0$ is preserved during the time evolution of the
system, using  (\ref{pbbTSmC0}) and
 (\ref{bTSNM}).
 As the next step  we analyze the requirement of the preservation of  the
constraints $\pi_i\approx 0$ during the time evolution of the system
\begin{eqnarray}
\partial_t\pi_i=\pb{\pi_i,H_E}=
- \left(\Omega_p\delta^k_i+\frac{\partial (\tD^{k}_{ \ j}\tn^j)}
{\partial
\tn^i}\right)\left(\mR_k-2m^2M_p^2\frac{\sqrt{g}}{\sqrt{\tx}}f_{km}
\tn^m \right)= 0 \ .
\end{eqnarray}
It turns out that the following matrix
\begin{equation}
\Omega_p \delta^k_i +\frac{\partial (\tD^k_{ \ j}\tn^j)} {\partial
\tn^i}=0 \
\end{equation}
cannot be solved for $\Omega_p$ and hence we have to demand an
existence of  following secondary constraints
\cite{Hassan:2011vm,Hassan:2011hr, Hassan:2011zd,Hassan:2011tf}
\begin{equation}
\mC_i\equiv \mR_i-\frac{2m^2M_p^2\sqrt{g}}{\sqrt{\tx}} f_{ij} \tn^j
\approx 0 \ .
\end{equation}
Finally we have to proceed to the analysis of the time development
of the constraint $\Sigma_p\approx 0$. Following
\cite{Kluson:2012zz} we simplify this constraint as follows.
   Using $\mC_i$ and $\Sigma_i$ we express $\tn^i$ as
a function of the  phase space variables $p_A,\phi^A$ and
$g_{ij},\pi^{ij}$ \cite{Kluson:2012zz}
\begin{equation}\label{tni3}
 \tn^i=-\frac{\partial_j\phi^Ap_A
f^{ji}}{\sqrt{p_A\partial_k\phi^A f^{kl}
\partial_l\phi^Bp_B +4m^4M_p^4 g}}+ \tilde{F}^{ij}\Sigma_j+\tilde{G}^{ij}\mC_j \
,
\end{equation}
where  $\tilde{F}^{ij},\tilde{G}^{ij}$ are phase space functions
whose explicit form is not needed for us.

Now using these results we find that the  constraint
 $\Sigma_p$ takes the form
\begin{eqnarray}
\Sigma_p
=\tSigma_p+H^i\Sigma_i+G^i\mC_i \ , \nonumber \\
\end{eqnarray}
where we introduced new independent constraint $\tSigma_p$
\begin{equation}\label{Sigmapfin}
\tSigma_p=4m^4M_p^4g+p_A\mG^{AB}p_B+
2T\sqrt{f}\sqrt{p_A\partial_i\phi^A
f^{ij}\partial_j\phi^Bp_B+4m^4M_p^4g}+T^2 f =0 \
\end{equation}
which is more complicated than in pure massive case due to the term
proportional to $T$. On the other hand we observe that in case when
$m=0$ this constraint takes the form $\tSigma_p= p_A\mG^{AB}p_B+
2T\sqrt{f}\sqrt{p_A\partial_i\phi^A f^{ij}\partial_j\phi^Bp_B}+T^2
f$ which means that $\tSigma_p$ is the linear combination of the
Hamiltonian and diffeomorphism constraints of the pure Galileon
action that is reviewed in appendix. In other words in the limit
$m\rightarrow 0$ the theory possesses eight first class constraints
$\mC_0,\mR_i,\mH_i,\mH_T$ that reflects  the fact that this theory
is invariant under two independent diffeomorphism.

Returning to the case $m\neq 0$ we define  the total Hamiltonian
with all constraints included
\begin{equation}\label{HTfin}
H_T=\int d^3\bx (N\mC_0+v_N\pi_N+v^i\pi_i+ \Omega_p \tSigma_p+
\Omega^i\tSigma_i+\Gamma^i\mC_i) \ .
\end{equation}
 Now we are ready to
analyze the stability of all constraints that appear in
(\ref{HTfin}).  First of all we find that $\pi_N\approx 0 $ is
automatically preserved while the preservation of the constraint
$\pi_i\approx 0$ gives
\begin{eqnarray}\label{parttpi}
\partial_t\pi_i&=&\pb{\pi_i,H_T}\approx
\int d^3\bx \Gamma^j(\bx)\pb{\pi_i,\mC_j(\bx)} = \nonumber \\
&=&-2m^2\Gamma^j\frac{1}{\sqrt{\tx}} (f_{ij}-f_{ik}\tn^k
f_{jl}\tn^l)\equiv -\triangle_{\pi_i,\mC_j}\Gamma^j \ .
\nonumber \\
\end{eqnarray}
By definition
\begin{eqnarray}
\det (f_{ij}-f_{ik}\tn^k f_{il}\tn^l)
=\tx \det f_{ij}\neq 0
 \nonumber \\
\end{eqnarray}
and hence the matrix $\triangle_{\pi_i,\mC_j}$  is non-singular.
Then the only solution of the equation (\ref{parttpi}) is
 $\Gamma^i=0$.

As the next step   we  perform the analysis of the stability of the
constraint $\tSigma_p$ . We introduce the smeared form of this
constraint
\begin{equation}
\bSigma(N)=\int d^3\bx N(\bx)\tSigma_p(\bx) \ .
\end{equation}
To proceed further we need following Poisson brackets
\begin{eqnarray}\label{paf}
\pb{p_A(\bx),f_{ij}(\by)}&=& -\partial_{y^i} \delta(\bx-\by)
\mG_{AB}\partial_{y^j}\phi^B(\by)-
\partial_{y^i}\phi^B(\by)\mG_{BA}\partial_{y^j}
\delta(\bx-\by) \ , \nonumber \\
\pb{p_A(\bx), f(\by)}&=&[ -\partial_{y^i} \delta(\bx-\by)
\mG_{AB}\partial_{y^j}\phi^B(\by)f^{ji}(\by)-
\partial_{y^i}\phi^B(\by)\mG_{BA}\partial_{y^j}
\delta(\bx-\by)f^{ji}(\by)] f(\by) \ , \nonumber \\
\pb{p_A(\bx),f^{ij}(\by)}&=& -f^{im}(\by) \pb{p_A(\bx), f_{mn}(\by)}
f^{nj}(\by)=
 \nonumber \\
 &=&f^{im}(\by)[\partial_{y^m}\delta(\bx-\by)\mG_{AB}\partial_{y^n}
 \phi^B(\by)+\partial_{y^m}\phi^B\mG_{BA}\partial_{y^n}
 \delta(\bx-\by)]f^{nj}(\by) \ .  \nonumber \\
\end{eqnarray}
With the help of these results  and after some calculations we
derive  following Poisson bracket
\begin{eqnarray}
\pb{\Sigma(N),\Sigma(M)}
&=&4T \int d^3\bx (N\partial_iM-M\partial_iN)
f^{ij}(\partial_j\phi^Ap_A)\frac{\sqrt{f}}{\sqrt{\bA}} \times \nonumber \\
&\times& (p_A\mG^{AB}p_B+T^2 f+4m^2M_p^2 g+2T\sqrt{f}\sqrt{\bA})= \nonumber \\
&=&4T\Sigma\left((N\partial_iM-M\partial_iN)f^{ij}(\partial_j\phi^Ap_A)\frac{\sqrt{f}}{\sqrt{\bA}}\right)
\ ,
\nonumber \\
\end{eqnarray}
where
\begin{equation}
\bA=p_A\partial_i\phi^A f^{ij}\partial_j\phi^Bp_B+4m^4M_p^4g \ .
\end{equation}
This is very important result that shows that the Poisson bracket
between $\tSigma_p$ vanishes on the constraint surface.

Finally we have to determine the Poisson bracket between $\Sigma(N)$
and $\bC(M)$ where
\begin{equation}
\bC(M)=\int d^3\bx M(\bx)\mC_0(\bx) \ .
\end{equation}
Using again (\ref{paf}) and after some calculations we find
following result
\footnote{Note that during  calculations we used the formula
\begin{eqnarray}
\frac{\delta ( \sqrt{\tx}\tD^k_{ \ k})}{\delta f_{ij}}
=\frac{\sqrt{\tx}}{2}\tD^j_{ \ p} f^{pi} -\frac{1}{\sqrt{\tx}} \tn^l
f_{lm}\frac{\delta( \tD^m_{ \
p}\tn^p)}{\delta f_{ij}} \ \nonumber \\
\end{eqnarray}
which follows  from (\ref{deftD}).}
 \begin{eqnarray}\label{SigmaC}
\pb{\Sigma(N),\bC(M)}&=&\int d^3\bx N M\Sigma^{II}+ \int d^3\bx
N\partial_iM \left[2T\frac{\sqrt{f}}{\sqrt{\bA}}
\partial_j\phi^Ap_A \frac{\delta (\tD^k_{ \ l}\tn^l)}{\delta
f_{ij}}\mC_k+ \right.\nonumber \\
&+& 2m^2M_p^2 T\frac{\sqrt{f}}{\sqrt{\bA}} \sqrt{g}\sqrt{\tx}
\tD^i_{ \
k} f^{kj}\Sigma_j - 2m^2M_p^2 T\sqrt{g}\sqrt{\tx} \tD^i_{ \ k} f^{kj}\mC_j+ \nonumber \\
&+& \left.
4\partial_j\phi^Ap_A\frac{\delta (\tD^k_{ \ l}\tn^l)}
{\delta f_{ij}}\mC_k +4m^2M_p^2\sqrt{g}\sqrt{\tx} \tD^i_{ \
j}f^{jk}\Sigma_k- 4m^2M_p^2
\sqrt{g}\sqrt{\tx}\tD^i_{ \ j}f^{jk}\mC_k\right] \ ,  \nonumber \\
\end{eqnarray}
where
\begin{eqnarray}
\Sigma_p^{II}&=& 4m^2M_p^2p_A\partial_i[\sqrt{g}\sqrt{\tx} \tD^i_{ \
p}f^{pj}\partial_j\phi^A]+ 4p_A\partial_i\left[\frac{\delta (\tD^k_{
\ l}\tn^l)}
{\delta f_{ij}}\mC_k\partial_j\phi^A\right]+ \nonumber \\
 &+&8m^4M_p^2\left(1+T\frac{\sqrt{f}}{\sqrt{\bA}}\right)
  \sqrt{g}g^{ij}\mG_{ijkl}\pi^{kl}+
4m^4M_p^4\left(1+\frac{\sqrt{f}}{\sqrt{\bA}}\right)
\left(2\partial_i[\tD^i_{ \ j}\tn^j]g+\partial_ig \tD^i_{ \
 j}\tn^j\right)+\nonumber \\
 &+&2m^2M_p^2 T\frac{\sqrt{f}}{\sqrt{\bA}}\partial_k\phi^Ap_A
 f^{kl}\partial_l\phi^B\partial_i
 [\sqrt{\tx}\sqrt{g}\tD^i_{ \ p}f^{pj}]+
 \nonumber \\
 &+&2T\frac{\sqrt{f}}{\sqrt{\bA}}
\partial_k\phi^Ap_Af^{kl}\partial_l\phi^B
\partial_i\left[\frac{\delta (\tD^m_{ \ n}\tn^n)}{\delta f_{ij}}
\mC_m\partial_j\phi_B\right] \nonumber \\
\end{eqnarray}
so that we see that (\ref{SigmaC}) vanishes on the constraint
surface up to the expression that we denote as $\tSigma_p^{II}$
\begin{eqnarray}
\tSigma^{II}_p&=& 2m^2M_p^2 \left(2p_A +
T\frac{\sqrt{f}}{\sqrt{\bA}}\partial_k\phi^Bp_B
 f^{kl}\partial_l\phi_A\right)\partial_i
 [\sqrt{\tx}\sqrt{g}\tD^i_{ \ p}f^{pj}\partial_j\phi^A]
+\nonumber \\
 &+&8m^4M_p^2\left(1+T\frac{\sqrt{f}}{\sqrt{\bA}} \right)
  \sqrt{g}g^{ij}\mG_{ijkl}\pi^{kl}+
4m^4M_p^4\left(1+\frac{\sqrt{f}}{\sqrt{\bA}}\right)
\left(2\partial_i[\tD^i_{ \ j}\tn^j]g+\partial_ig \tD^i_{ \
 j}\tn^j\right) \ . \nonumber \\
\end{eqnarray}
Now we are ready to proceed to the analysis of the requirement of
the preservation of the constraint $\tSigma_p$
\begin{eqnarray}\label{timeSigmap}
\partial_t \tSigma_p=\pb{\tSigma_p,H_T}\approx
\int d^3\bx N(\bx)\pb{\Sigma_p,\mC_0(\bx)}\approx \int d^3\bx
N(\bx)\tSigma_p^{II}(\bx) \ .
\nonumber \\
\end{eqnarray}
From this result we see  that   the constraint $\tSigma_p\approx 0$
is preserved  during the time evolution of the system on condition
when either $N=0$ or when $\tSigma^{II}_p=0$. Note that we should
interpreted $N$ as the Lagrange multiplier so that it is possible to
demand that $N=0$ on condition when $\tSigma^{II}_p\neq 0$ on the
whole phase space. It seems to us that  such a condition is too
strong so that it is more natural to demand that
$\tSigma_p^{II}\approx 0 $ and $N\neq 0$. In other words
$\tSigma_p^{II}\approx 0$ is the new secondary constraint.

It is convenient to  have constraint $\tSigma_p^{II}$ independent on
$\tn^i$. This can be easily done when we use (\ref{tni3}) and insert
it
 into the explicit form of $\tD^i_{ \ j}$
\cite{Hassan:2011vm,Hassan:2011hr, Hassan:2011zd,Hassan:2011tf}
\begin{eqnarray}
\tD^i_{ \ j}&=&\sqrt{g^{im}f_{mn}Q^n_{ \ p}}(Q^{-1})^p_{ \ j} \ ,
\nonumber \\
 Q^m_{ \ p}&=&\tx \delta^m_{ \ p}+\tn^m\tn^n f_{np} \ , \quad
(Q^{-1})^p_{ \ j}=\frac{1}{\tx}(\delta^m_p-\tn^p\tn^m f_{mj}) \
\nonumber \\
\end{eqnarray}
so that  we find
\begin{eqnarray}
Q^m_p&=&\frac{1}{A+4M_p^4 m^4 g}(4m^4M_p^4g\delta^m_{ \ p}+
\partial_j\phi^Ap_A f^{jm}\partial_p \phi^Bp_B) \ , \nonumber \\
(Q^{-1})^m_p&=&\frac{A+4M_p^4 m^4 g}{4m^4 M_p^4 g}\left(\delta^m_{ \
p}-\frac{1}{A+4m^4 M_p^4 g}
\partial_j\phi^Ap_A f^{jm}\partial_p \phi^Bp_B\right) \  \nonumber \\
\end{eqnarray}
up to terms proportional to the constraints $\mC_i,\Sigma_i$. With
the help of these results it is easy to formulate $\tSigma_p^{II}$
as a constraint that does not depend on $\tn^i$ (Again up to the
terms proportional to $\Sigma_i,\mC_i$). This fact simplifies
further analysis considerably since now the Poisson brackets between
$\tSigma_p^{II}$ and $\pi_i$ are zero.

In summary we have following collection of constraints:
$\pi_N\approx 0 \ , \pi_i\approx 0 \ ,  \mC_0\approx 0, \mC_i\approx
0, \tSigma_i\approx 0 , \tSigma_p\approx 0 , \tSigma^{II}_p\approx
0$. The dynamics of these constraints is governed by the total
Hamiltonian
\begin{equation}
H_T=\int d^3\bx (N\mC_0+v_N\pi_N+v^i\pi_i+\Omega_p\tSigma_p+
\Omega_p^{II}\tSigma_p^{II}+\Omega^i\tSigma_i+\Gamma^i\mC_i) \ .
\end{equation}
As the final step we have to analyze  the preservation of all
constraints, following
 \cite{Kluson:2012zz}. The case of $\pi_N\approx 0$ is trivial.
 In case of  $\pi_i\approx 0$ we obtain
\begin{eqnarray}\label{parttpii2}
\partial_i\pi_i(\bx)&=&\pb{\pi_i(\bx),H_T}=
\int d^3\by( \Gamma^j(\by)\pb{\pi_i(\bx),\mC_j(\by)}
+\Omega_p^{II}(\by) \pb{\pi_i(\bx),\tSigma_p^{II}(\by)})=
 \ \nonumber \\
&=& \Gamma^j\triangle_{\pi_i,\mC_j}(\bx)=0
\nonumber \\
\end{eqnarray}
due to the crucial fact that we used the formulation when
$\tSigma_p^{II}$ does not depend on $\tn^i$. Then as we argued above
the only solution of the equation  is $\Gamma^i=0$. Now the time
development of $\mC_i$ is given by the equation
\begin{eqnarray}\label{parttCi}
\partial_t \mC_i(\bx)&=&\pb{\mC_i(\bx),H_T}
\approx \nonumber \\
&\approx & \int d^3\bx \left(N(\by)\pb{\mC_i(\bx),\mC_0(\by)}+
v^j(\by)
\pb{\mC_i(\bx),\pi_j(\by)}+\right.\nonumber \\
&+&\left.\Omega_p(\by)\pb{\mC_i(\bx),\tSigma_p(\by)}+
\Omega_p^{II}(\by)
\pb{\mC_i(\bx),\tSigma_p^{II}(\by)}\right) \nonumber \\
\end{eqnarray}
and the time development of the constraint $\tSigma_p$ is governed
by the equation
\begin{eqnarray}
\partial_t\tSigma_p(\bx)=
\pb{\tSigma_p(\bx),H_T}\approx \int d^3\bx \Omega_p^{II}(\by)
\pb{\tSigma_p(\bx),\tSigma^{II}_p(\by)} \ .
\nonumber \\
\end{eqnarray}
As follows from the explicit form of the constraint $\tSigma_p^{II}$
we  see that $\pb{\tSigma_p^{II}(\bx),\tSigma_p(\by)}$ is non-zero
and proportional also to the  higher order  derivatives of the delta
functions.  As a consequence we find that the only solution of the
equation above is $\Omega_p^{II}=0$. Further we analyze the time
evolution of the constraint $\tSigma_p^{II}$
\begin{eqnarray}
\partial_t\tSigma_p^{II}(\bx)&=&\pb{\tSigma_p^{II}(\bx),H_T}=
\nonumber \\
&=&\int d^3\bx \left(N(\by)\pb{\tSigma_p^{II}(\bx),\mC_0(\by)}+
\Omega_p(\by) \pb{\tSigma^{II}_p(\bx),\tSigma_p(\by)}
\right)=0 \ .  \nonumber \\
\end{eqnarray}
 Now
from the last equation we obtain $\Omega_p$ as a function of the
phase space variables and $N$, at least in principle. Then inserting
this result into the equation for the preservation of $\mC_i$
(\ref{parttCi}) we determine $v^j$ as functions of the phase space
variables.
 Finally note also that the
constraint $\mC_0$ is automatically preserved due to the fact that
$\Gamma^i=\Omega_p^{II}=0$ and also the fact that
$\pb{\mC_0(\bx),\mC_0(\by)}\approx 0$ as was shown in
\cite{Hassan:2011ea}.

 In
summary we obtain following picture. We have five  first class
constraints $\pi_N\approx 0\ , \mC_0\approx 0 \ , \tSigma_i\approx
0$ together with eight  second class constraints $\pi_i\approx 0 \ ,
\mC_i\approx 0$ and $\tSigma_p\approx 0 \ , \tSigma_p^{II}\approx
0$. The constraints $\pi_i\approx 0$ together with $\mC_i\approx 0$
can be solved for $\pi_i$ and $\tn^i$. Then the constraint
$\tSigma_p$ can be solved for $g=\det g_{ij}$ in terms of $p_A$ and
$\phi^A$ while the constraint $\tSigma_p^{II}$ is the single
constraint on $g_{ij}$ and $\pi^{ij}$. Altogether these two
constraints are responsible for the elimination of the
Boulware-Deser ghost. As a result we have $10$ gravitational degrees
of freedom ,$2D+2$ scalars degrees of freedom together with $4$
first class constraints $\mC_0\approx 0 \ , \tSigma_i\approx 0$.
Then we find that the number of physical degrees of freedom is
$10+2(D+1)-8=10+2(D-3)$ which corresponds to the number of physical
degrees of freedom of the massive gravity coupled with $D-3$ scalar
Galileon fields. In other words we have shown that this specific
model of the minimal dRGT gravity coupled with tadpole Galileon
field is ghost free.
\\
 \noindent {\bf
Acknowledgement:}
 This work   was
supported by the Grant agency of the Czech republic under the grant
P201/12/G028. \vskip 5mm

\begin{appendix}
\section{Hamiltonian Analysis of
Galileon Tadpole Term} In this section we briefly review the
Hamiltonian analysis of the  Galileon tadpole term
\begin{equation}\label{Sgalltadapp}
S_{gal}=-T\int d^4x\sqrt{-\det f_{\mu\nu}} \ .
\end{equation}
 The momentum conjugate to $\phi^A$ takes the form
\begin{eqnarray}\label{paapp}
p_A
=-T\mG_{AB}\partial_\mu\phi^B (f^{-1})^{\mu 0} \sqrt{-\det
f_{\mu\nu}} \ .
\nonumber \\
\end{eqnarray}
Taking the square of given expression  we find following primary
constraint
\begin{eqnarray}
\mH_T=p_A\mG^{AB}p_B+ T^2f\approx 0 \ .
\nonumber \\
\end{eqnarray}
On the other hand when we multiply (\ref{paapp}) with
$\partial_i\phi^A$ we obtain following $3$ primary constraints
\begin{equation}
\mH_i=\partial_i\phi^Ap_A
\approx 0
\end{equation}
Introducing the smeared forms of these constraints $H_T(N)=\int
d^3\bx N\mH_T$ and $H_S(N^i)=\int d^3\bx N^i\mH_i$ we easily find
the Poisson brackets
\begin{eqnarray}
\pb{H_T(N),H_T(M)}&=&2T^2H_S((\partial_iMN-\partial_iNM)f^{ij} \det
f) \ ,
\nonumber \\
\pb{H_S((N^i),H_T(M)}&=&H_T((N^i\partial_iM-\partial_iN^i)) \ ,
\nonumber \\
\pb{H_S(N^i),H_S(M^j)}&=&H_S((N^i\partial_iM^j-M^i\partial_iN^j)) \
\nonumber \\
\end{eqnarray}
that coincide with the Poisson brackets calculated for example in
\cite{Bengtsson:2000xa}. We see that $\mH_T,\mH_i$ are the first
class constraints which  is a consequence of diffeomorphism
invariance of given theory.
\end{appendix}


\end{document}